\documentclass[11pt]{article}
\usepackage{amssymb}
\begin{document}
\title{ $A^{0}Z^{0}$ associated production at the Large Hadron Collider
in the minimal supersymmetric standard model \footnote{Supported
by National Natural Science Foundation of China.}} \vspace{3mm}
\author{{ Yin Jun$^{2}$, Ma Wen-Gan$^{1,2}$, Zhang Ren-You$^{2}$, and Hou Hong-Sheng$^{2}$}\\
{\small $^{1}$ CCAST (World Laboratory), P.O.Box 8730, Beijing
100080, P.R.China}\\
{\small $^{2}$ Department of Modern Physics, University of Science and Technology}\\
{\small of China (USTC), Hefei, Anhui 230027, P.R.China}}
\date{}
\maketitle \vskip 12mm
\begin{abstract}
We investigate in detail the $A^{0}Z^{0}$ associated production
process $pp \rightarrow A^0Z^0+X$ within the framework of the
minimal supersymmetric standard model (MSSM) at the CERN Large
Hadron Collider (LHC), considering both contributions from the
Drell-Yan and gluon fusion subprocesses. We focus on the
deviations from the general two-Higgs-doublet model (2HDM)
arising in the MSSM. We also discuss the contributions of the two
$A^{0}Z^{0}$ associated production subprocesses in the MSSM at
the LHC, and analyse the dependences of the total cross section
on neutral CP-odd Higgs boson mass $m_A$ and $\tan\beta$ in the
mSUGRA scenario. We find that the contribution from loop mediated
gluon fusion subprocess can be competitive with that from the
Drell-Yan subprocess in some parameter space.
\end{abstract}
 \vfill \eject
\baselineskip=0.36in
\renewcommand{\theequation}{\arabic{section}.\arabic{equation}}
\renewcommand{\thesection}{\Roman{section}}
\newcommand{\nb}{\nonumber}
\makeatletter      
\@addtoreset{equation}{section}
\makeatother       
\section{Introduction}
\par
The minimal standard model(MSM) \cite{s1} \cite{s2} has been
proved by all precise experimental data that the MSM is a very
successful model of particle physics. But until now the symmetric
breaking structure of the electroweak interactions has not yet
been directly explored experimentally. So the exploration of the
SM Higgs boson is a major goal of the present and future
colliders. As we know, any enlargement of the Higgs sector beyond
the single $SU(2)_{L}$ Higgs doublet of the MSM necessarily
introduces other neutral Higgs bosons and charged Higgs bosons.
Like the general two-Higgs-doublet model(2HDM), the minimal
supersymmetric standard model (MSSM) \cite{s3} \cite{haber}
requires the introduction of two Higgs doublets in order to
preserve supersymmetry. These two Higgs doublets predict some more
elementary Higgs bosons: one CP-even neutral Higgs boson($H^0$),
one CP-odd neutral Higgs boson($A^0$) and two charged Higgs
bosons($H^{\pm}$), which are absent in the MSM. Any experimental
discovery of these non-SM-like Higgs bosons will be the direct
verification of these extended versions of the Higgs sector.
Therefore, the study of various production mechanisms of the
non-SM-like Higgs bosons at the present and future colliders is
well motivated.
\par
Searching for the non-SM-like Higgs bosons and studying their
properties at the future multi-TeV hadron colliders, such as the
CERN Large Hadron Collider (LHC), are possible as expected by
supersymmetric (SUSY) theory \cite{s3} \cite{susy}. The gluon
fusion mechanism $g g \to \phi(\phi=h^0,H^0,A^0)$ provides the
dominant production mechanism of neutral Higgs bosons at the LHC
in the entire relevant mass range up to about 1 TeV for the small
and moderate values of $\tan\beta$ in the MSSM \cite{Gunion-a-1}.
The heavy neutral Higgs boson can be also produced in pair
($A^0A^0, A^0h^0, A^0 H^0$) at the LHC, if it is kinematically
allowed \cite{Kniehl-a-a}. Studying the process of a heavy
charged Higgs boson associated with $W$ boson is another
attractive way in searching for the $H^{\pm}$ bosons, because the
$W^{\pm}$-boson's leptonic decay may be used as a spectacular
trigger. The calculations of the heavy $H^{\pm}$ production
associated with $W^{\mp}$ boson at a future electron-positron
collider can be found in Refs. \cite{eehw} \cite{eehwa}
\cite{eehwb} \cite{eehwc} \cite{eehwd}. The complete calculations
of the $H^{\pm} W^\mp$ associated production at hadron colliders
both in the 2HDM and the MSSM are given in Refs. \cite{Dicus}
\cite{Kniehl} \cite{Kniehl1} \cite{zhou} \cite{Hollik-a-1}. 
Analogously, $A^0 Z^0$
associated production would also be an efficient way in searching
for the heavy neutral CP-odd Higgs boson $A^0$. Although the
$A^0$ boson can be produced in pair at future colliders
\cite{Kniehl-a-a} \cite{ppaa}, the $A^0 Z^0$
associated production will be the kinematically favored mechanism
to produce $A^0$ Higgs boson for the heavy $A^0$ Higgs boson. And
again the leptonic decay of $Z^0$ maybe benefit for triggering
the $Z^0 A^0$ associated production events. The calculations of
the $Z^0 A^0$ associated production at a electron-positron
collider were presented in Refs. \cite{eeza1}
\cite{eeza2}, at a muon collider in Ref. \cite{Akeroyd-a-1} and
at a photon collider in Ref. \cite{Gounaris-a-1}, respectively.
And Chung Kao gave the calculation of 
$A^0Z^0$ associated production via $g g$ fusion including only 
quark loop diagrams at the SSC \cite{Kao}.
\par
In this paper we concentrate on studying the $A^0Z^0$ associated
production at the LHC in the MSSM, considering both subprocesses
$q\bar q\rightarrow A^0Z^0$ and $gg \rightarrow A^0Z^0$.
In the
calculation of the loop mediated process $pp \to gg \rightarrow
A^0Z^0$, we compare and discuss the cross sections in the general
two-Higgs-doublet model (2HDM) and the MSSM. In section II, we
present the calculation of the processes $pp \to q\bar
q\rightarrow A^0Z^0$ and $pp \to gg \rightarrow A^0Z^0$.
Numerical results and discussion are given in section III. There
we use the MSSM parameters constrained within the minimal
supergravity (mSUGRA) scenario \cite{msugra}. Finally, a short
summary is given.
\par
\section{Cross Section Calculation}
In our calculation we use the t'Hooft-Feynman gauge and adopt the
dimension regularization scheme in the general 2HDM and the
dimensional reduction (DR) scheme \cite{DR scheme} in the MSSM.
In the loop diagram calculation we adopted the definitions of
one-loop integral functions in reference \cite{s13}. The
numerical calculation of the vector and tensor loop integral
functions can be traced back to scalar loop integrals as shown in
the reference\cite{s14}. The Feynman diagrams and the relevant
amplitudes are created by FeynArts package automatically
\cite{denner-a-1}. The numerical calculation of the loop
integrals are implemented by using Mathematica programs.

\subsection{Calculation of the subprocess $q\bar q \to
A^0Z^0+X$}
\par
We denote the $A^0Z^0$ associated production via Drell-Yan
subprocess as
\begin{equation}
q(p_1)+\bar{q}(p_2) \to Z^0(k_1)+ A^0(k_2),
\end{equation}
\par
Due to the feature of the Yukawa coupling that the coupling
strength between quarks and Higgs boson is in proportion to the
correspondent quark mass, the cross sections of subprocesses
$q\bar q\rightarrow A^0Z^0(q=u,d,s,c)$ should be much smaller
than those of the subprocesses $t\bar t(b \bar b)\rightarrow
A^0Z^0$. Considering the fact that the luminosity of top
(anti-top) quark is much lower than that of bottom (anti-bottom)
quark from a proton, we conclude that the cross section of the
process $pp \rightarrow q\bar q \rightarrow A^0Z^0+X$ is
approximately equal to the cross section of $pp \rightarrow b\bar
b \rightarrow A^0Z^0+X$. Therefore, in this paper we consider only
the contributions from the $pp \to b\bar b \to A^0Z^0+X$ process.
\par
The Feynman diagrams of the subprocess $b\bar b \rightarrow
A^0Z^0$ at the lowest order are depicted in Fig.1. The
differential cross section of the subprocess $b\bar b\rightarrow
A^0Z^0$ can be expressed as
\begin{equation}\label{eq:xsec0}
d {\hat \sigma_{b\bar b}}=  dP_{2f} \, \frac{1}{12} \sum_{spin}
 |A_{(a)}(\hat s,\hat t, \hat u)+ A_{(b)}(\hat s,\hat t, \hat u)|^2,
\end{equation}
where the summation is taken over the spins of the initial and
final states, and $dP_{2f}$ denotes the two-particle phase space
element. The factor $1/12$ in above equation comes from the
averaging over the spins and the colors of the incoming partons.
The matrix element $A_{(a)}$ represents the amplitude of the
$h^0$($H^0$) exchanging s-channel diagrams(shown in Fig.1(a)),
$A_{(b)}$ corresponds to the amplitude of u- and t-channel
diagrams (shown in Fig.1(b)). The Mandelstam kinematical
variables are defined as
\begin{equation}
\hat{s} = (p_1 + p_2)^2, \quad \hat t = (p_1-k_1)^2, \quad \hat u
= (p_1-k_2)^2, \label{eq:mandel}
\end{equation}
By using the relevant Feynman rules, we obtain the explicit
expressions of these amplitudes:
\begin{eqnarray}
A_{(a)}(\hat{s},\hat{t}, \hat{u})
       &=& -\frac{(4\pi\alpha)m_b}{4s_{W}^2c_{W}^2}\cos(\beta-\alpha)
           \frac{\sin\alpha}{\cos\beta}\epsilon^\mu(k_1)
       \left[\bar{v} (p_2) u(p_1)\right]
       \frac{(k_1+p_1+p_2)^{\mu}}{\hat{s}-m_{h}^2+m_h\Gamma_h i} \nb \\
       & & -\frac{(4\pi\alpha_s)m_b}{4s_{w}^2c_{W}^2}\sin(\beta-\alpha)
           \frac{\cos\alpha}{\cos\beta}\epsilon^\mu(k_1)
       \left[\bar{v} (p_2) u(p_1)\right]
       \frac{(k_1+p_1+p_2)^{\mu}}{\hat{s}-m_{H}^2+m_H\Gamma_H i} \\
A_{(b)}(\hat{s},\hat{t}, \hat{u})
       &=& -\frac{(4\pi\alpha)m_b}{2m_W
           s_{W}^2c_{W}}\tan\beta\epsilon^\mu(k_1)\frac{1}
           {\hat{t}-m_b^2}\left[\bar{v}
           (p_2)\gamma^5(m_b-\rlap/{k_1}+\rlap/{p_1})\gamma^\mu(\frac{s_{W}^2}{3}-\frac{P_L}{2})
           u(p_1)\right] \nb \\
      & & -\frac{(4\pi\alpha)m_b}{2m_W
          s_{W}^2c_W}\tan\beta\epsilon^\mu(k_1)\frac{1}
          {\hat{u}-m_b^2}\left[\bar{v}
          (p_2)\gamma^5(m_b+\rlap/{k_1}-\rlap/{p_2})\gamma^\mu(\frac{s_{W}^2}{3}-\frac{P_L}{2})
          u(p_1)\right] \nb
\end{eqnarray}
where $m_b$ and $m_W$ represent the masses of bottom quark and W
boson, respectively.
\par
\subsection{Calculation of the subprocess $gg \to
A^0Z^0 +X$}
\par
We denote the $A^0Z^0$ associated production process via gluon
fusions as
\begin{equation}
g(p_1,\alpha)+g(p_2,\beta) \rightarrow Z^0(k_1)+ A^0(k_2),
\end{equation}
where $\alpha$, $\beta$ are the color indices of initial gluons.
As the subprocess $gg \rightarrow Z^0A^0$ is loop-induced, the
one-loop order calculation can be simply carried out by summing
all unrenormalized reducible and irreducible one-loop diagrams
and the results will be finite and gauge invariant. We denote
$\sigma^{2HDM}_{gg}$ and $\sigma^{MSSM}_{gg}$ as the cross
sections in the framework of the general 2HDM and the MSSM,
respectively. The former is contributed by the Feynman diagrams
involving only the quark loop diagrams(shown in Fig.2) and the
latter involves the contributions of both the quark and squark
loop diagrams(shown in Fig.2-3). The possible corresponding
Feynman diagrams created by exchanging the initial gluons or the
two final states, should be also included in Fig.2 and Fig.3 and
involved in our calculation .
\par
We can see that each Feynman diagram in Fig.2 and Fig.3 contains
one interacting vertex between (s)quarks and a Higgs boson. Due to
the feature of the Yukawa coupling as we mentioned above, we can
consider only the diagrams which involve the third generation
(s)quark in the calculation of the subprocess $gg \to A^0Z^0$.
The cross sections of the subprocess $gg \rightarrow A^0Z^0$ in
the general 2HDM and the MSSM can be expressed respectively as
\begin{eqnarray}
d {\hat \sigma^{2HDM}_{gg}} &=& dP_{2f} \frac{1}{256} \sum
 |A^{(2)}_{(a)}(\hat s,\hat t, \hat u)+ A^{(2)}_{(b)}(\hat s,\hat t,
 \hat u)+A^{(2)}_{(c)}(\hat s,\hat t, \hat u)|^2 \nb \\
d {\hat \sigma^{MSSM}_{gg}} &=& dP_{2f} \frac{1}{256} \sum
 |A^{(2)}_{(a)}(\hat s,\hat t, \hat u)+ A^{(2)}_{(b)}(\hat s,\hat t,
 \hat u)+A^{(2)}_{(c)}(\hat s,\hat t, \hat u) \nb \\
 & & + A^{(3)}_{(a)}(\hat s,\hat t, \hat u)+A^{(3)}_{(b)}(\hat s,\hat t,
  \hat u)\cdots +A^{(3)}_{(g)}(\hat s,\hat t, \hat u)|^2
\end{eqnarray}
where the summation is taken over the spins and colors of the
initial and final states, and $dP_{2f}$ denotes the two-particle
phase space element. $A^{(i)}_{(j)}$ represents the amplitude of
the diagram of Fig.i(j). The factor $1/256$ results from the
averaging over the spins and the colors of the incoming partons.

\subsection{Cross section of $p p \rightarrow A^0 Z^0+X$ process at the LHC}
\par
With the cross sections of the related subprocesses, the cross
section of parent process $p p \rightarrow A^0 Z^0+X$ at the
proton-proton collider LHC can be obtained by doing the following
integration,
\begin{eqnarray}
\label{integration}
\sigma_{ij}= \int_{(m_Z+m_A)^2/ s} ^{1} d \tau \frac{d%
{\cal L}_{ij}}{d \tau} \hat{\sigma}_{ij}(\hat{s}=\tau s)
\end{eqnarray}
where
\begin{eqnarray}
\frac{d{\cal L}_{ij}}{d\tau}=\frac{1}{1+\delta_{ij}}\int_{\tau}^{1} \frac{%
dx_1}{x_1} \left\{\left[
f_{i/p}(i,x_1,Q^2)f_{j/p}(j,\frac{\tau}{x_1},Q^2) \right] +\left[
f_{j/p}(j,x_1,Q^2)f_{i/p}(i,\frac{\tau}{x_1},Q^2) \right]\right\}
\end{eqnarray}
In Eq.(\ref{integration}) $\sqrt{s}$ and $\sqrt{\hat{s}}$ are the
colliding proton-proton and parton-parton c.m.s. energies
respectively. The notation $\sigma_{ij}$ represents the cross
section of the parent process $pp \rightarrow ij \rightarrow
A^0Z^0+X$. $d{\cal L}_{ij}/d \tau$ is the luminosity of incoming
partons where $i,~j$ can be $b,~\bar{b}$ and $g$, $\tau =
x_1~x_2$. $m_Z$ and $m_A$ represent the masses of $Z^0$ boson and
$A^0$ Higgs boson. The definitions of $x_1$ and $x_2$ can be
found in Ref.\cite{x1x2}. In our calculation, we adopt the CTEQ5
parton distribution function \cite{function} and take the
factorization scale Q to be $\sqrt{\hat{s}}$. The
Eq.(\ref{integration}) can be rewritten as
\begin{eqnarray}
\sigma_{ij}= \int_{m_Z+m_A} ^{\sqrt{s}} d \sqrt{\hat{s}}
 \hat{\sigma}_{ij }(\hat{s})H_{ij}(\hat{s})
\end{eqnarray}
where
\begin{eqnarray}
H_{ij}(\hat{s})=\frac{1}{1+\delta_{ij}}\int_{\frac{\hat {s}}{s}}^{1} \frac{%
  2dx_1\sqrt{\hat{s}}}{ x_1s} \left\{\left[
f_{i/p}(i,x_1,Q^2)f_{j/p}(j,\frac{\hat{s}}{x_1s},Q^2) \right]
+(i \leftrightarrow j) \right\}
\end{eqnarray}
When $ij=gg$, $\sigma^{2HDM}_{gg}$ represents the cross section
of the parent process $pp \rightarrow gg \rightarrow A^0Z^0+X$
contributed only by quark loop diagrams shown in Fig.2, while
$\sigma^{MSSM}_{gg}$ represents the cross section contributed by
both quark and squark loop diagrams shown in Fig.2 and Fig.3. In
the next section we shall take different input data sets to
demonstrate the production rates of the parent process
$pp\rightarrow A^0Z^0+X$. The numerical results of
$\sigma^{2HDM}_{gg}$ and $\sigma^{MSSM}_{gg}$ would show the
importance of squark loop diagrams. The total cross section of
$pp \to A^0Z^0+X$ at proton-proton collider should be the
summation of $\sigma_{b\bar b}$ and $\sigma_{gg}$. Quantitatively
comparing the $\sigma_{b\bar b}$ with $\sigma_{gg}$ will help us
to know in which part of the parameter space the contribution of
gluon-gluon fusion process is dominant.

\par
\section{Numerical result and discussion}
\subsection{Input parameters}
\par
In the numerical calculation, we take the SM parameters as: $m_t
=174.3~$GeV, $m_b =4.2~$GeV, $m_Z = 91.187~$GeV, $\Gamma_Z =
2.49~$GeV \cite{s12}, and take the supersymmetric parameters being
constrained within the minimal supergravity (mSUGRA) scenario
\cite{msugra}. In this scenario, only five sypersymmetric
parameters should be inputed, namely $M_{1/2}$, $M_0$, $A_0$,
sign of $\mu$ and $\tan\beta$, where $M_{1/2}$, $M_0$ and $A_0$
are the universal gaugino mass, scalar mass at GUT scale and the
trilinear soft breaking parameter in the superpotential terms,
respectively. In this work, we take $M_{1/2}$=120 GeV, $A_0$=300
GeV and $\mu>0$. $M_0$ is obtained quantitatively from the input
$m_{A}$ value. All other MSSM parameters are determined in the
mSUGRA scenario by using program package ISAJET 7.44. In this
program, the renormalization group equations (RGE's) \cite{RGE}
are run from the weak scale $m_Z$ up to the GUT scale, taking all
thresholds into account in order to get the low energy scenario
from the mSUGRA. It uses two loop RGE's only for the gauge
couplings and the one-loop RGE's for the other supersymmetric
parameters. The GUT scale boundary conditions are imposed and the
RGE's are run back to $m_Z$, again taking threshold into account.
\par
Here we give some comments about the choice of the decay width
values of CP-even neutral Higgs bosons $h^0$ and $H^0$. We know
that some of the Feynman diagrams (shown in Fig.1-3) have
s-channel $h^0$ and $H^0$ propagators, which have analytical
expressions respectively as
\begin{equation}
\frac{1} {\hat{s}-m_{h}^2+ i m_h\Gamma_h}=\frac{\hat{s}- i
m_{h}^2-
m_h\Gamma_h} {(\hat{s}-m_{h}^2)^2+m_h^2\Gamma_h^2 },\\
\end{equation}
\begin{equation}
\frac{1} {\hat{s}-m_{H}^2+ i m_H\Gamma_H}=\frac{\hat{s}-m_{H}^2-
i m_H\Gamma_H} {(\hat{s}-m_{H}^2)^2+m_H^2\Gamma_H^2 },\\
\end{equation}
It is clear that the cross sections of the subprocess should
related to the decay widths of $h^0$ and $H^0$. In this work the
input parameter $m_A$ is taken in the range of 200 GeV to 650
GeV. Then we have the following constraints in this parameter
space,
\begin{equation}
m_H\approx m_A, ~~ \quad m_h<150~ {\rm GeV},
\end{equation}
and by using the package HDECAY\cite{HDecay} in the MSSM, we find
\begin{equation}
\Gamma_H,~\Gamma_h<10~ {\rm GeV}.
\end{equation}
Because $\sqrt{\hat s}\geq m_A+m_Z$, we get $(\hat{s}-m_{H}^2)^2
\gg m_H^2\Gamma_H^2$. The propagator of $H^0$ boson can be
expressed approximately as
\begin{equation}
\frac{\hat{s}-m_{H}^2- i m_H\Gamma_H}
{(\hat{s}-m_{H}^2)^2+m_H^2\Gamma_H^2 }\approx \frac{1}
{(\hat{s}-m_{H}^2) }
\end{equation}
It is obvious that the bigger the $\sqrt{\hat s}$ is, the less
sensitive the cross section to the decay widths of neutral Higgs
bosons $H^0$ and $h^0$ is. Therefore, we choose
$\Gamma_H=\Gamma_h=10$~GeV in our numerical calculations.
Actually, our final numerical result of the cross section of the
process $pp \to A^0Z^0+X$ at the LHC, shows also that it is not
sensitive to the choice of these two decay widths.

\par
\subsection{Discussion and analysis}
\par
The figures in Fig.4, Fig.5 and Fig.6, show the cross sections
(or differential cross sections) of the process $pp \to gg \to
A^0Z^0+X$ at the LHC as the functions of the CP-odd Higgs boson
$A^0$ mass, the ratio of the vacuum expectation values
$\tan{\beta}$ and the transverse momentum $p_T$, respectively.
The curves of the cross sections (or differential cross sections)
involving the contributions from quark loop diagrams (in the
general 2HDM) and quark+squark loop diagrams (in the MSSM) are
depicted separately on these figures for comparison of the cross
sections in these two models. And in these three figures the
full-lines are for $\sigma^{2HDM}_{gg}$ (or
$d\sigma^{2HDM}_{gg}/dp_T$), the dotted-lines are for
$\sigma^{MSSM}_{gg}$ (or $d\sigma^{MSSM}_{gg}/dp_T$).
\par
Figure 4 shows the relationship between the cross section of the
parent process $pp \to gg \to A^0Z^0+X$ and $m_A$ with the
colliding energy $\sqrt{s}=14~$TeV. The input mSUGRA parameters
are set to be the typical values mentioned in the last subsection
(i.e. $M_{1/2}$=120 GeV, $A_0$=300 GeV and $\mu>0$. $M_0$ is
obtained quantitatively from the $m_{A}$ value), and
$\tan{\beta}$=2, 7 and 32, respectively. From this figure, we
find that in some parameter space the scalar quark contributions
can enhance the cross section obviously, that is to say
$\sigma^{MSSM}_{gg}>\sigma^{2HDM}_{gg}$, while in other regions,
we have $\sigma^{MSSM}_{gg}<\sigma^{2HDM}_{gg}$. The figure shows
that when we have small and moderate $\tan\beta$ values, the
scalar quark loop contribution to the $A^0Z^0$ associated
production at the LHC is most obvious. We shall also see later
from Fig.7 that when $\tan {\beta}$ has small or moderate value,
the contributions from gluon fusion subprocess is dominant.
Therefore, it is possible to use the experimental measurement of
the $A^0Z^0$ associated production at the LHC to disentangle the
MSSM from the general 2HDM in these parameter space regions.
\par
In Fig.5, the cross section of the parent process $pp \to gg \to
A^0Z^0+X $ at the LHC versus $\tan{\beta}$ is plotted. The values
of the neutral CP-odd Higgs boson $A^0$ mass are set to be 200
GeV, 400 GeV and 600 GeV, respectively. From the figure, we also
find that scalar quark contributions can either enhance or
suppress the cross section of the parent process as shown in
Fig.4. Fig.5 together with Fig.4, show that when the value of
$m_A$ is greater than 400 GeV and $\tan{\beta}<8$, the
contribution from the scalar quark loop diagrams increases with
the decrement of $\tan{\beta}$. In Fig.5 the two curves for
$m_A=400~{\rm GeV}$ and $600~{\rm GeV}$ demonstrate that when
$\tan{\beta}<8$, the scalar quark contribution suppresses the
cross section, which means $\sigma^{MSSM}_{gg}<
\sigma^{2HDM}_{gg}$, and while the scalar quark contribution
enhances the cross section when $\tan{\beta}>20$. These features
can be also seen from Fig.4. The curve for $\tan{\beta}=2$ in
Fig.4, demonstrates that $\sigma^{MSSM}_{gg}$ is about two third
of $\sigma^{2HDM}_{gg}$ quantitatively, while the curve for
$\tan{\beta}=32$ shows $\sigma^{MSSM}_{gg}>\sigma^{2HDM}_{gg}$.
\par
Fig.6 displays the differential cross section $d\sigma/dp_T$ of
the process $pp \to gg \to A^0Z^0 $ at the LHC versus transverse
momentum $p_T$ with $\sqrt{s}=14~TeV$ and the pseudo-rapidity
being in the range of $|\eta|<2$. The $A^0$ mass is set to be 350
GeV, and $\tan{\beta}$ is taken as 2, 7 and 32, respectively. We
find that for $\tan{\beta}=2$, the scalar quark contribution
suppresses the differential cross section $d\sigma/dp_T$. For
$\tan{\beta}=32$ and $p_T>100~$GeV, the scalar quark contribution
enhances the differential cross section. But for $\tan\beta=7$,
the scalar quark contribution can either enhance or suppress the
differential cross section in different $p_T$ regions.
\par
In Fig.7, Fig.8 and Fig.9, we plot the the cross sections of the
process $pp \to b \bar{b} \to A^0Z^0+X$ ($\sigma^{(DY)}$), and the
process $pp \to A^0Z^0+X$ contributions from both Drell-Yan and
gluon-gluon fusion subprocesses in the constrained MSSM
($\sigma^{(T)}=\sigma^{(DY)} + \sigma^{MSSM}_{gg}$), as the
functions of the CP-odd Higgs boson $A^0$ mass, the ratio of the
vacuum expectation values $\tan{\beta}$ and the transverse
momentum $p_T$, respectively. In these three figures, the
full-lines are for the cross sections or differential cross
sections of the process $pp \to b \bar{b} \to A^0Z^0+X$ via
Drell-Yang subprocess, the dotted-lines are for the process $pp
\to A^0Z^0+X$ via both Drell-Yang and gluon fusion subprocesses.
With the comparison between the $\sigma^{(T)}$ (or
$d\sigma^{(T)}/dp_T$) and $\sigma^{(DY)}$ (or
$d\sigma^{(DY)}/dp_T$), we can know in which parameter space in
the constrained MSSM, the contribution from the loop mediated
subprocess $gg \to A^0Z^0 $ is important.
\par
Fig.7 displays the cross sections of $pp \to b\bar b \to A^0Z^0
+X$ and $pp \to A^0Z^0+X $ at proton-proton colliders versus the
mass of $A^0$ with $\sqrt{s}=14~$TeV. We choose $\tan{\beta}$=2,
7 and 32, respectively. From Fig.7, we find that in the region of
$\tan{\beta}\le 7$, the contribution of gluon-gluon fusion
subprocess in the MSSM enhances the cross section, especially
when $\tan\beta=2$ the contribution of the gluon-gluon fusion
subprocess is about $80\%$ of the total cross section
$\sigma^{(T)}$. In fact, the gluon-gluon fusion subprocess is the
most important $A^0Z^0$ associated production mechanism in this
parameter space. From the figure we see also that when
$\tan{\beta}$=32 the difference between $\sigma^{(DY)}$ and
$\sigma^{(T)}$ is very small, it means that the contribution of
the gluon-gluon fusion subprocess is negligible in this parameter
space.
\par
The cross sections of $pp \to b\bar b \to A^0Z^0+X $ and $pp \to
A^0Z^0+X$ at the LHC as the functions of $\tan{\beta}$ with
$\sqrt{s}=14~TeV$ are shown in Fig.8. The mass of Higgs boson
$A^0$ is taken as 200 GeV, 400 GeV and 600 GeV, respectively.
From this figure we can find also that gluon-gluon fusion
subprocess enhances the cross section of the $A^0Z^0$ associated
production at the LHC, and will become a very important production
mechanism when $\tan{\beta}<10$. In the region of
$\tan{\beta}>10$, the cross section of the $A^0Z^0$ associated
production at the LHC are in the range of $1-10^2$ fb. Even the
$\sigma^{(T)}$ can reach 300 fb when $\tan{\beta}=48$ and
$m_A=200~$GeV. So the $A^0Z^0$ associated production process may
be easily observed experimentally if $\tan{\beta}$ is large
enough.
\par
Fig.9 displays the differential cross sections ($d\sigma/dp_T$) of
$pp \to A^0Z^0+X $ and $pp \to b\bar b \to A^0Z^0+X $ at the LHC
as the functions of the transverse momentum $p_T$ with the
pseudo-rapidity being in the range of $|\eta|<2$. We choose
$m_A=350$ GeV, and take $\tan{\beta}$=2, 7 and 32, respectively.
From this figure we can see that at high $p_T$ region, when
$\tan{\beta}\le 7$, the difference between $d\sigma^{(DY)}/dp_T$
and $d\sigma^{(T)}/dp_T$ is obvious, even when $\tan{\beta}$=2,
the $d\sigma^{(DY)}/dp_T$ can be less than $1\%$ of
$d\sigma^{(T)}/dp_T$, which means that the contribution from the
$pp \to gg \to A^0Z^0+X $ process is dominant in this parameter
space. But when $\tan\beta=32$, the contribution to the total
differential cross sections ($d\sigma^{(T)}/dp_T$) is mainly from
the Drell-Yan $A^0Z^0$ associated production subprocess, and the
contribution from gluon fusion subprocess is negligible.

\par
\section{ Summary}
\par
In this paper, we studied the neutral CP-odd Higgs boson $A^0$
production with the association of $Z^0$ gauge boson via both
Drell-Yan and gluon-gluon fusion subprocesses in the constrained
MSSM at the CERN LHC. Numerical analysis of their production rates
is carried out with some typical parameter sets in the mSUGRA
scenario. Our results show that the cross section in the MSSM is
clearly enhanced by the gluon-gluon fusion subprocess in the
parameter space with small or moderate $\tan{\beta}$ value, and
we should consider the gluon-gluon fusion subprocess in this
parameter space in the calculation of the $A^0Z^0$ associated
production at the LHC. We compared above results of the process
$pp \to gg \to A^0Z^0 +X$ in the MSSM with those in the general
two-Higgs-doublet model (2HDM), where the cross section of
subprocess $gg \to A^0Z^0+X$ is contributed only by quark loop
diagrams. We find that the contributions from the scalar quark
loops in the MSSM can either enhance or suppress the cross
section obviously and cannot be neglected in some parameter
space. The results show also that the $A^0Z^0$ associated
production at the LHC is strongly related to the parameters
$\tan{\beta}$ and the mass of $A^0$. The total cross section
increases with increment of $\tan{\beta}$, and decreases with
increment of $m_A$.
\par
\noindent{\large\bf Acknowledgments:} This work was supported in
part by the National Natural Science Foundation of China and a
grant from the Education Ministry of China .

\vskip 10mm \vskip 10mm
\begin{flushleft} {\bf Figure Captions} \end{flushleft}
\par
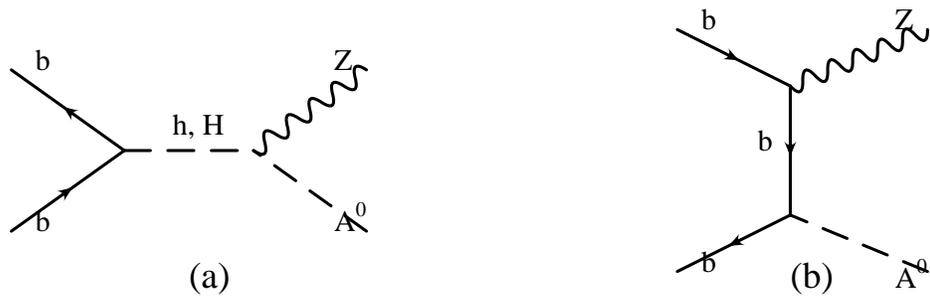
{\bf Fig.1} The relevant Feynman diagrams for the subprocess $b
\bar{b} \to A^0Z^0 $ in the MSSM at the tree-level: (a) s-channel
diagrams. (b) u- and t-channel diagrams. Note that Fig.1(b)
includes the diagram created by exchanging two final states.
\par
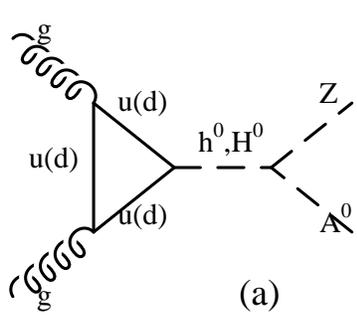
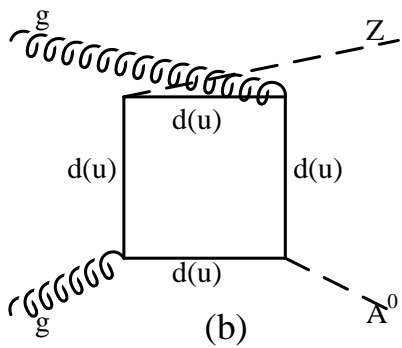
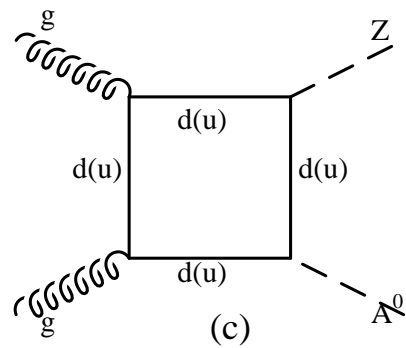
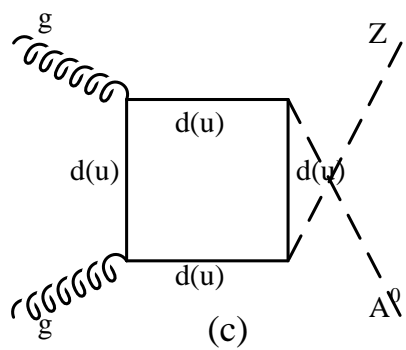
{\bf Fig.2} The relevant Feynman diagrams for the subprocess $gg
\to Z^0 A^0$ at the one loop-level (including only the quark loop
diagrams).
\par
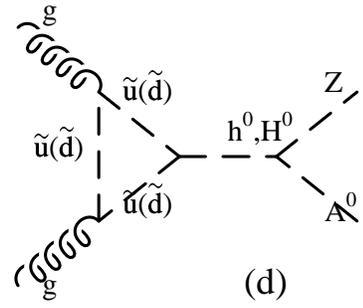
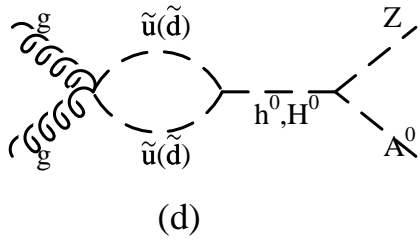
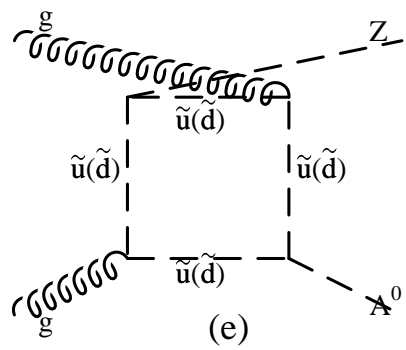
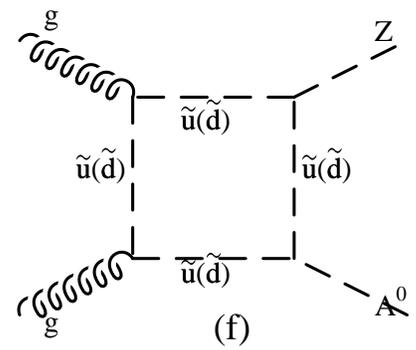
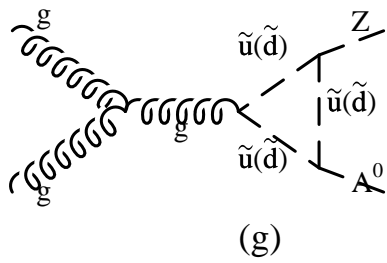
{\bf Fig.3} The relevant Feynman diagrams for the subprocess $gg
\to A^0Z^0 $ at the one loop-level (including only the scalar
quark loop diagrams).
\par
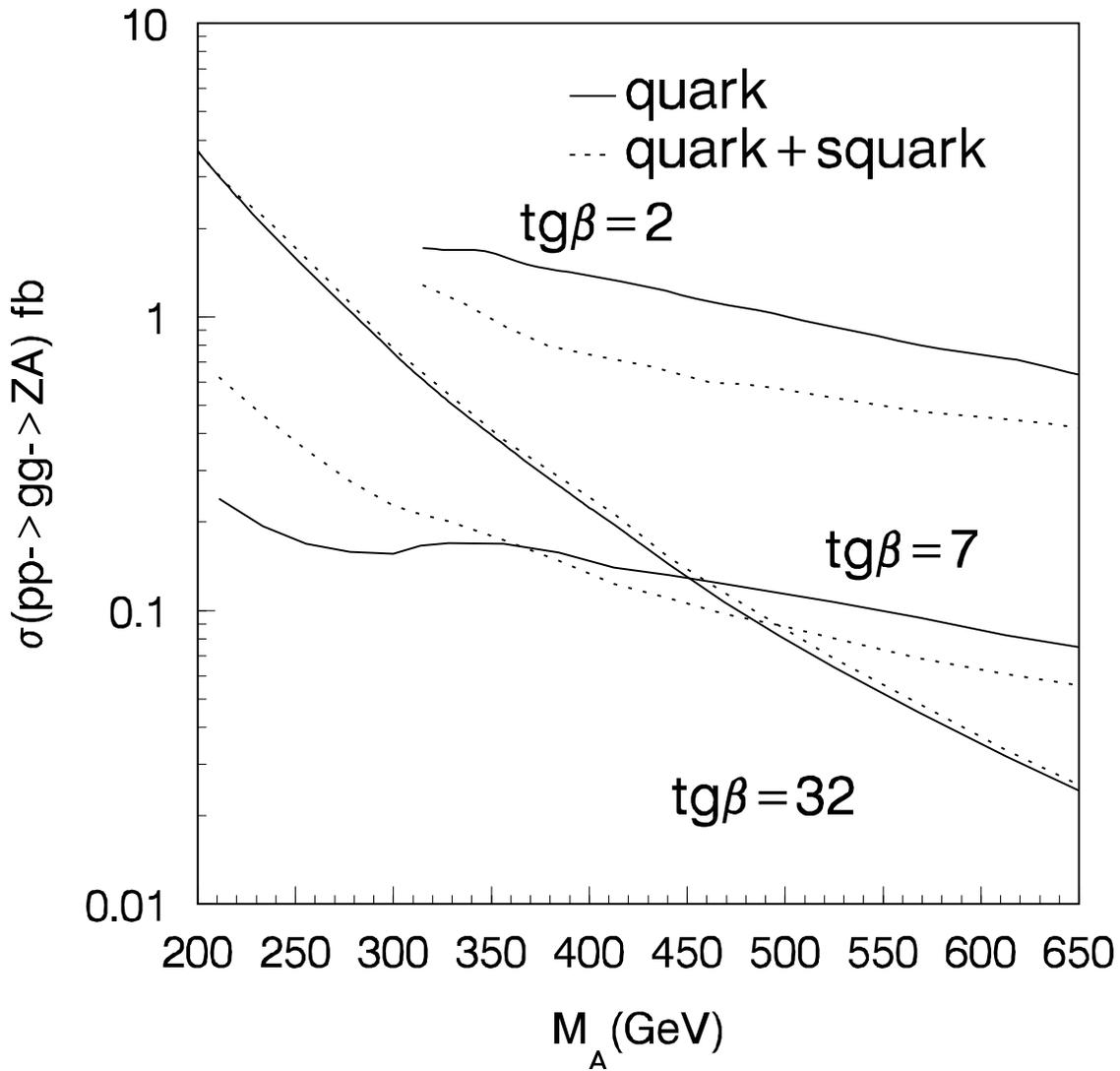
{\bf Fig.4} The cross sections $\sigma^{2HDM}_{gg}$ and
$\sigma^{MSSM}_{gg}$ of the process $pp \to gg \to A^0Z^0+X$, as
the functions of the mass of Higgs boson $A^0$. The input
parameter $\tan{\beta}$ is taken as 2, 7 and 32, respectively.
\par
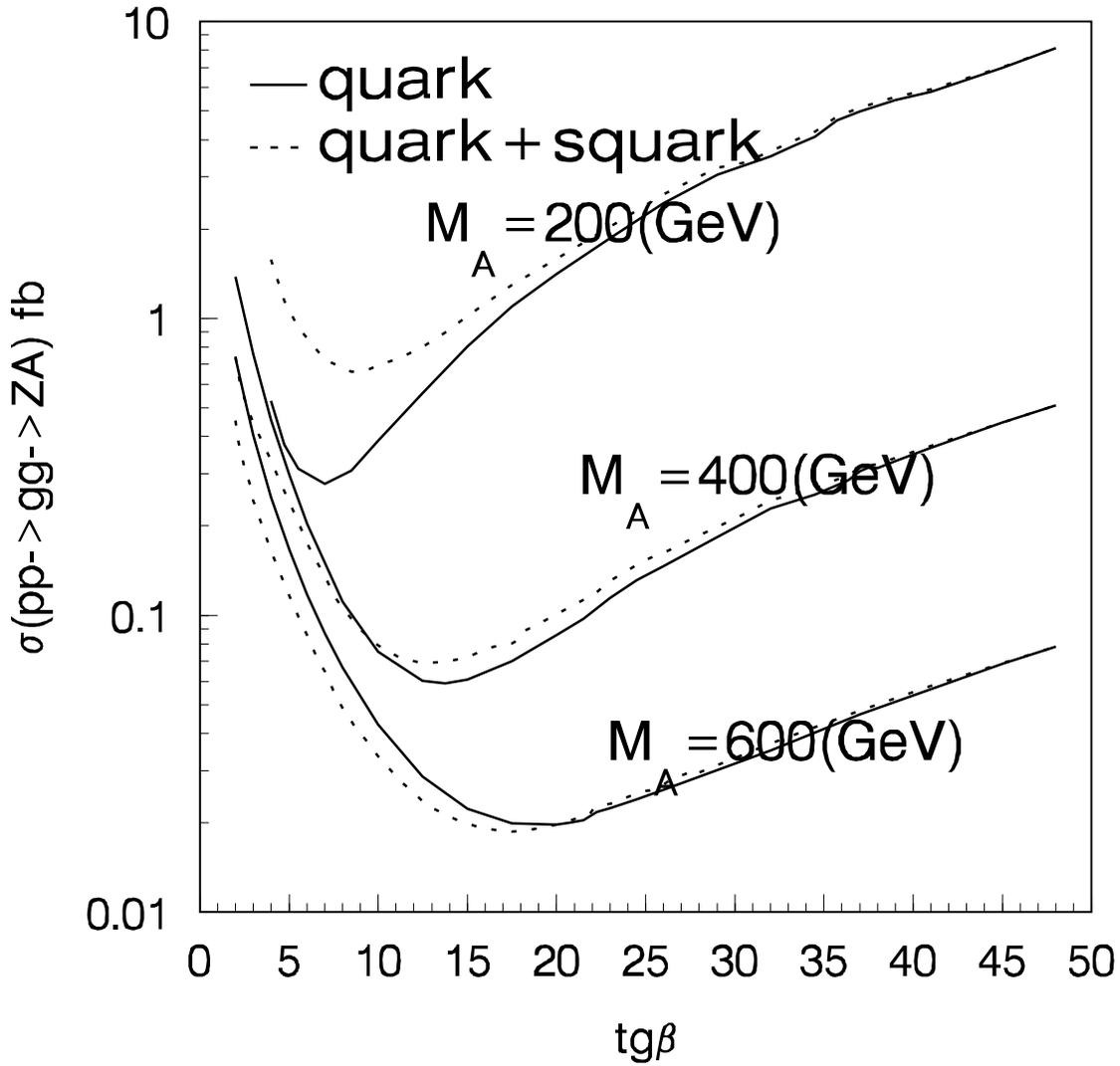
{\bf Fig.5} The cross sections $\sigma^{2HDM}_{gg}$ and
$\sigma^{MSSM}_{gg}$ of the process $pp \to gg \to A^0Z^0+X $, as
the functions of $\tan\beta$. The mass of Higgs boson $A^0$ is
taken as 200 GeV, 400 GeV and 600 GeV, respectively.
\par
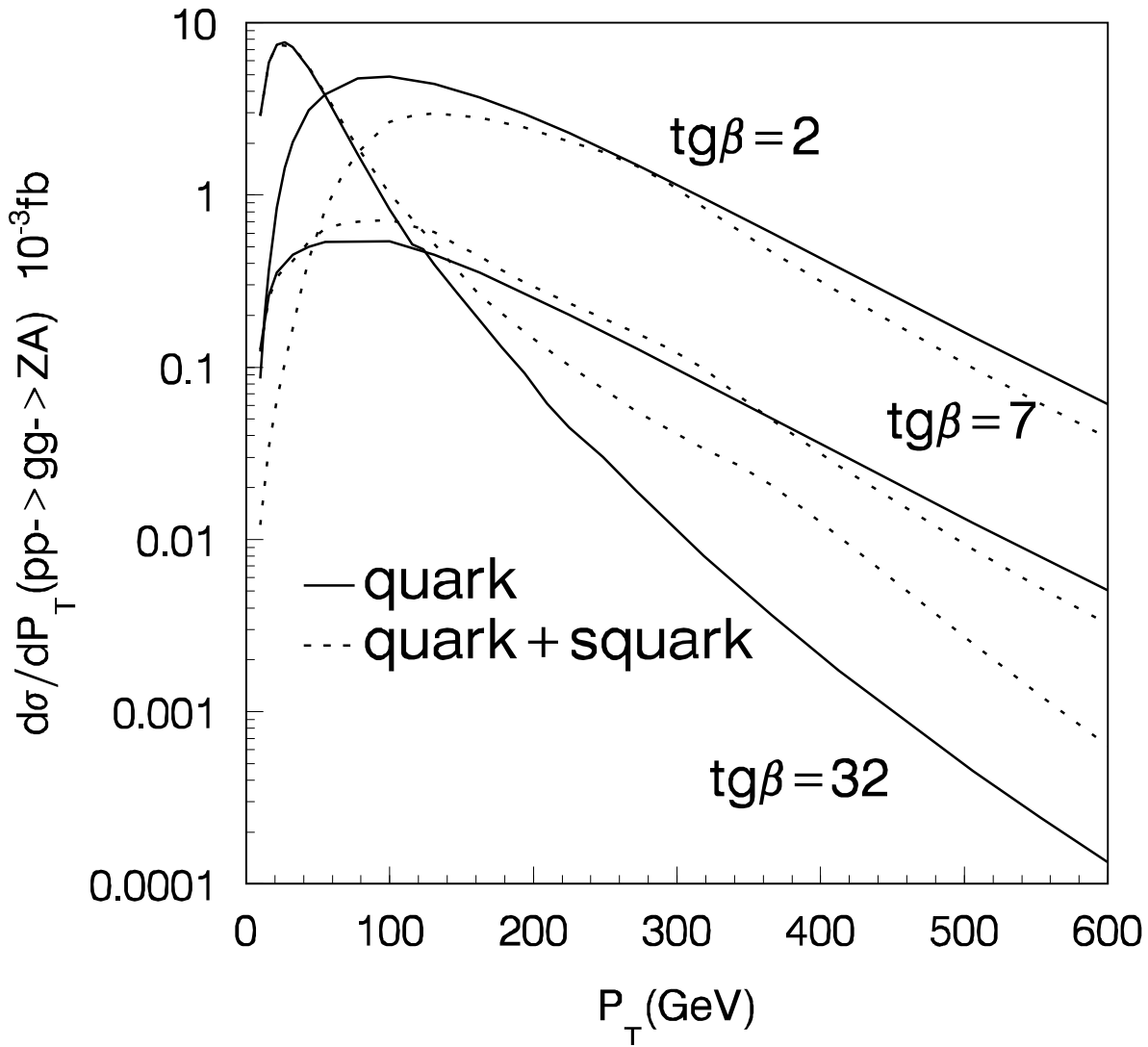
{\bf Fig.6} The differential cross sections
$d\sigma_{gg}^{2HDM}/dp_T$ and $d\sigma_{gg}^{MSSM}/dp_T$ of the
process $pp \to gg \to A^0Z^0+X $, as the functions of the
transverse momentum $p_T$ in the mSUGRA scenario at the LHC with
$\sqrt{s}=14~$TeV, $m_{A}=350~$GeV and the pseudo-rapidity being
in the range of $|\eta|<2$. The ratio of the vacuum expectation
values $\tan\beta$ is taken as 2, 7 and 32, respectively.
\par
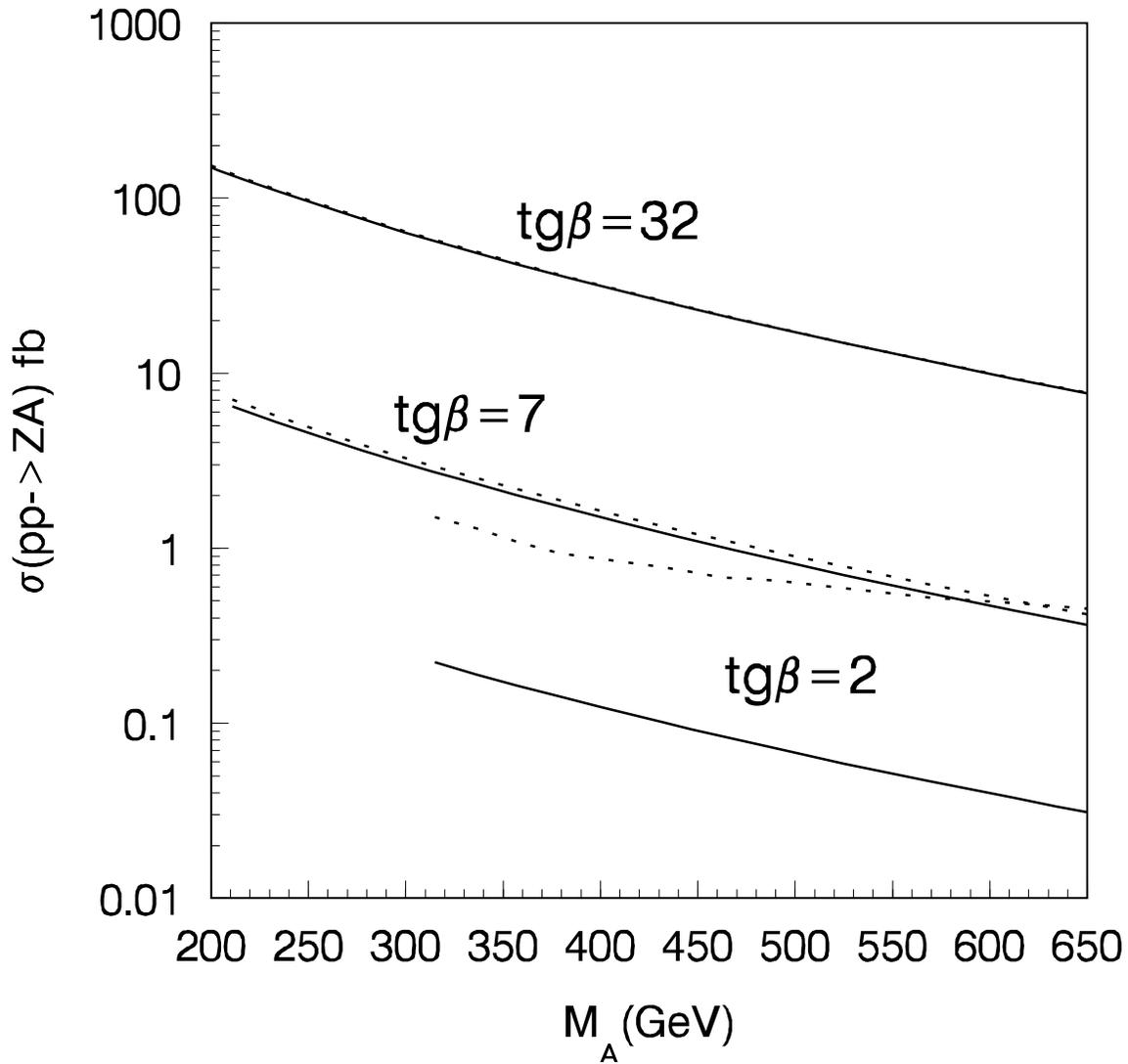
{\bf Fig.7} The cross sections $\sigma^{(DY)}$ and $\sigma^{(T)}$
of the process $pp \to A^0Z^0+X $ as the functions of the mass of
Higgs boson $A^0$,  The ratio of the vacuum expectation values
$\tan\beta$ is taken as 2, 7 and 32, respectively.
\par
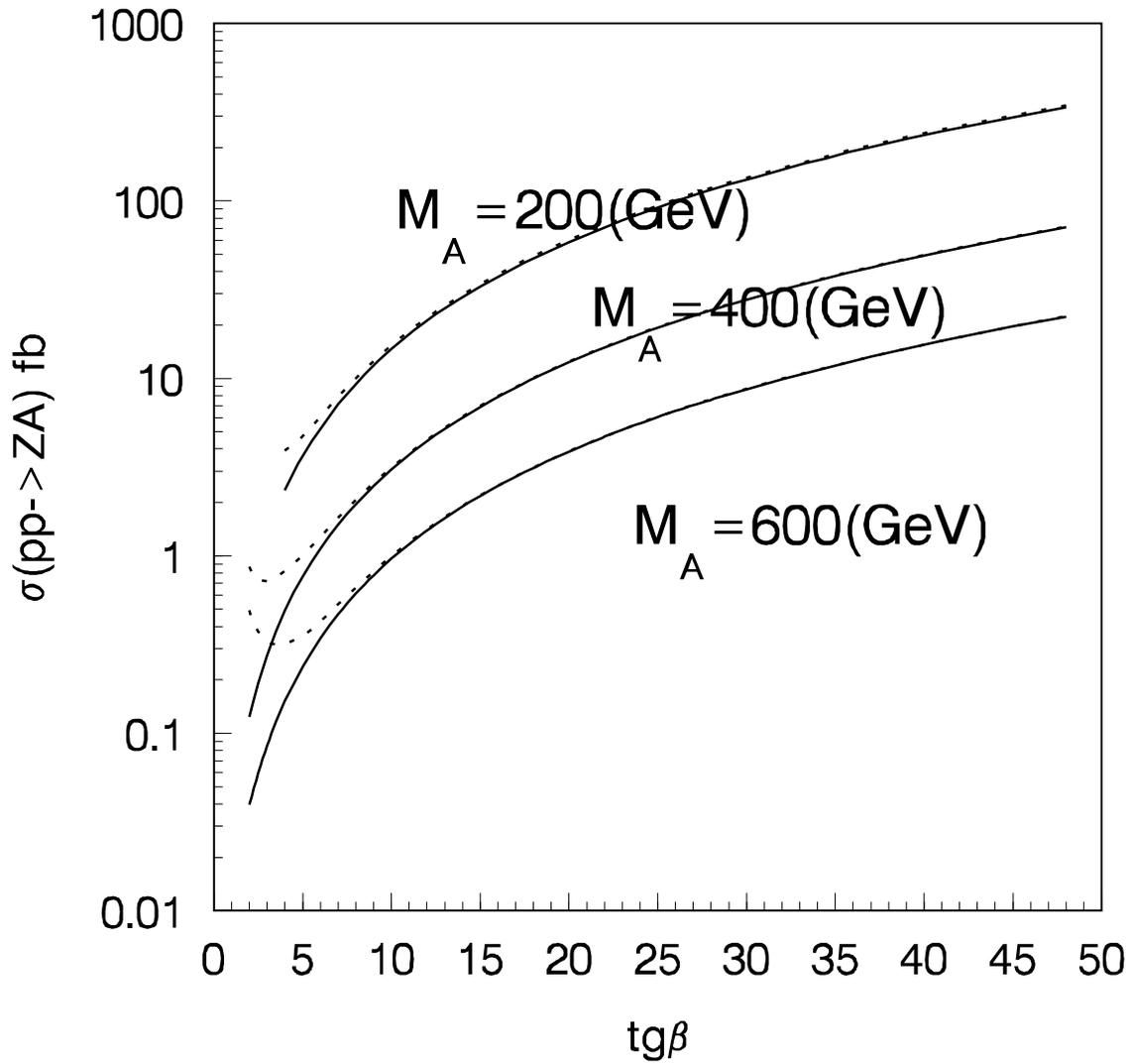
{\bf Fig.8} The cross sections $\sigma^{(DY)}$ and $\sigma^{(T)}$
of the process $pp \to A^0Z^0+X $ as the functions of
$\tan\beta$. The mass of Higgs boson $A^0$ is taken as 200 GeV,
400 GeV and 600 GeV, respectively.
\par
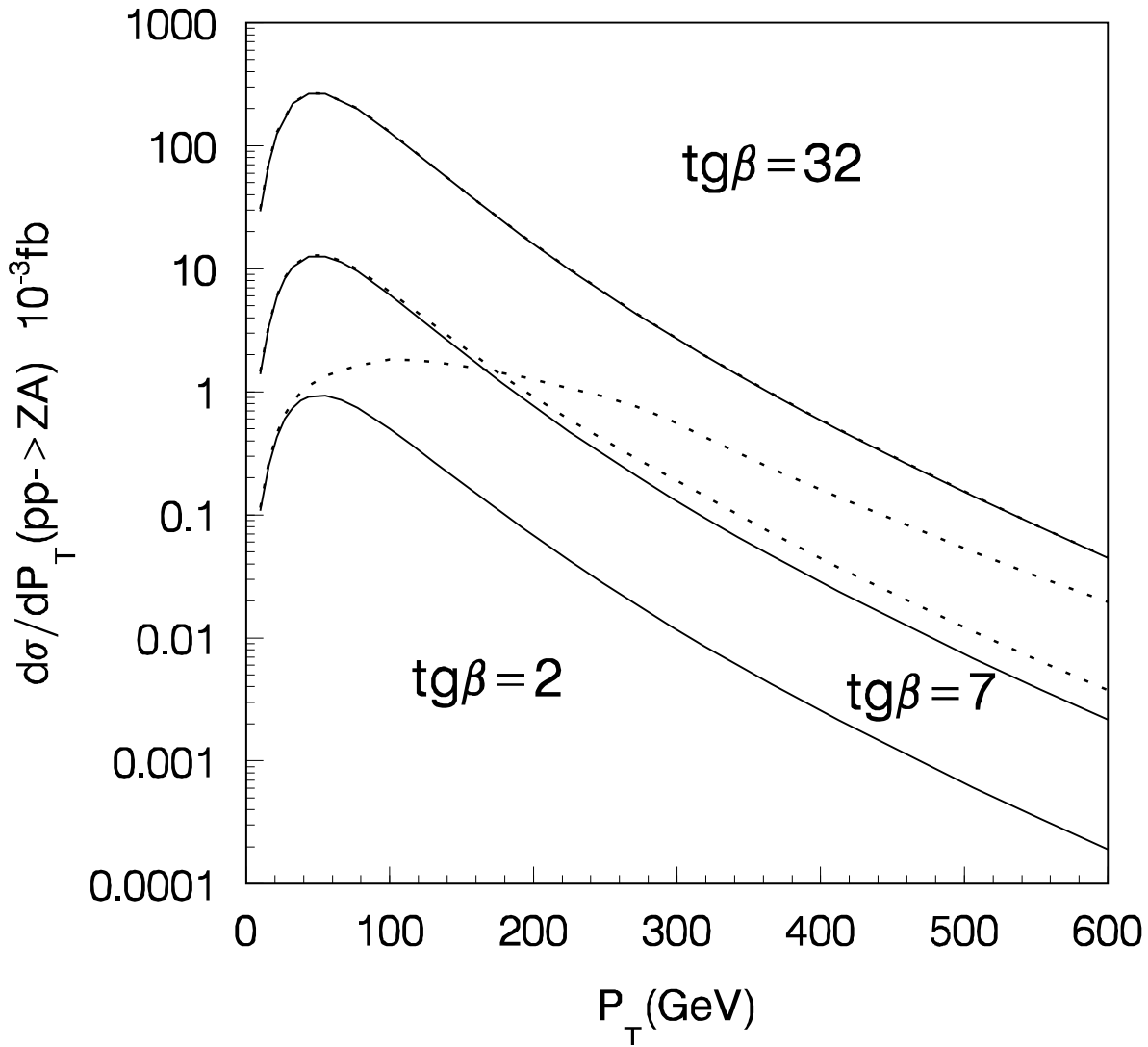
{\bf Fig.9} The differential cross sections $d\sigma^{(DY)}/dp_T$
and $d\sigma^{(T)}/dp_T$ of the process $pp \to A^0Z^0+X $, as the
functions of the transverse momentum $p_T$ in the mSUGRA scenario
with $\sqrt{s}=14~$TeV, $m_{A}=350$~GeV and the pseudo-rapidity
being in the range of $|\eta|<2$. The ratio of the vacuum
expectation values $\tan\beta$ is taken as 2, 7 and 32,
respectively.
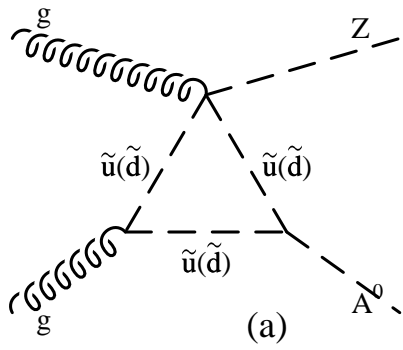
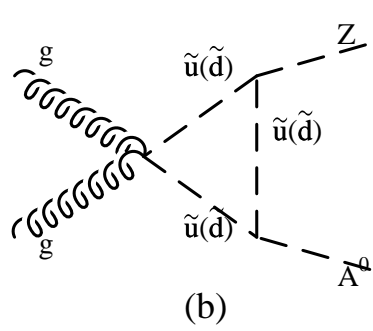
\end{document}